# Reflectance measurement of two-dimensional photonic crystal nanocavities with embedded quantum dots

**Wolfgang C. Stumpf,\* Takashi Asano, Takanori Kojima, Masayuki Fujita, Yoshinori Tanaka, and Susumu Noda**

*Department of Electronic Science and Engineering, Kyoto University, Kyotodaigaku-katsura,*

*Nishikyo-ku, Kyoto 615-8510, Japan*

*\*Corresponding author: wolfgang@qoe.kuee.kyoto-u.ac.jp*

# Abstract

The spectra of two-dimensional photonic crystal slab nanocavities with embedded InAs quantum dots are measured by photoluminescence and reflectance. In comparing the spectra taken by these two different methods, consistency with the nanocavities' resonant wavelengths is found. Furthermore, it is shown that the reflectance method can measure both active and passive cavities. *Q* factors of nanocavities, whose resonant wavelengths range from 1280 to 1620 nm, are measured by the reflectance method in cross polarization. Experimentally, *Q* factors decrease for longer wavelengths and the intensity, reflected by the nanocavities on resonance, becomes minimal around 1370 nm. The trend of the *Q* factors is explained by the change in the slab thickness relative to the resonant wavelength, showing a good agreement between theory and experiment. The trend of reflected intensity by the nanocavities on resonance can be understood as effects that originate from the photonic crystal slab and the underlying air-cladding thicknesses. In addition to three-dimensional finite-difference time-domain calculations, an analytical model is introduced that is able to reproduce the wavelength dependence of the reflected intensity observed in the experiment.

PACS numbers: 42.70.Qs, 42.82.Et, 42.50.Pq, 85.35.Be

# Introduction

Photonic nanocavities with a high-quality factor $Q$, small mode volume $V_m$, [1,2] and embedded light emitters have attracted much attention in recent years. They can be used to probe, investigate, and manipulate the fundamental physics of light-matter interaction, such as spontaneous emission, [3,4] and strong coupling, [5,6] and other effects, classified as cavity quantum electrodynamics (CQED). [7] The system of a nanocavity with embedded light emitters (active nanocavity) could potentially fit a broad range of applications, such as devices for highly efficient photon generation [8] and devices for quantum information processing systems. [9] Generally speaking, a nanocavity's $Q$ factor is an important parameter for the manipulation and control of emitter-cavity interaction. Most commonly, the $Q$ factor is estimated from the photoluminescence (PL) spectrum at emitter absorption saturation below the lasing threshold. [10] However, experiments suggest that in PL mutual interactions between emitter and cavity yield complicated effects [6,11,12] that make $Q$ factor evaluation difficult. [13] Alternatively, the $Q$ factor can be measured using a coupled waveguide [14] or a fragile near-field probe (such as a tapered fiber or a fiber tip). [15,16] Nevertheless, these methods require either special measurement geometry or sophisticated modifications of the setup. On the contrary, in the reflectance method, which was developed recently, probe light is introduced directly to the nanocavity from free space and can be readily applied. Thus far, reflection spectra of two-dimensional (2D) photonic crystal (PC) microcavities have been used to demonstrate slow group velocity of light, [17] polarization sensitivity, [18] and the probing of modes by resonant scattering of laser pulses have been reported. [19] Recently, we used a similar technique to investigate the $Q$ factors and the effects on $Q$ from embedded InAs quantum dots (QDs) in 2D-PC nanocavities. [20] Also, other groups have reported on QDs to control the nanocavity's reflectivity [21] or used CQED effects for QD spectroscopy. [22]

In this paper, we aim to consolidate the basics of the reflectance measurement of 2D-PC slab nanocavities. Nanocavities with embedded InAs QDs (Ref. 23) were prepared for a wide range of resonant wavelengths. The nanocavities were measured by both methods PL and reflectance, where the behavior of a nanocavity changes gradually from active to passive as the resonant wavelength is detuned from the QDs' emission wavelength. The consistency of both methods is confirmed for active nanocavities and the reflectance measurement is extended to passive nanocavities. It is observed that the thickness of the PC slab, relative to the resonant wavelength and the thickness of the lower air cladding play important roles in the nanocavity's reflection characteristics.

*Sample fabrication method*

The substrate was grown by molecular-beam epitaxy on a semi-insulating GaAs wafer. At 600 °C, after 200 nm of GaAs buffer layer, a 1200 nm thick $Al_{0.7}Ga_{0.3}As$ sacrificial layer and then a 100 nm GaAs PC layer were grown. Then the substrate temperature was decreased to 520 °C, the arsenic pressure lowered to 5.6 x $10^{-7}$ mbar, and an amount of InAs equivalent to 2.6 monolayers (MLs) was supplied at a growth rate of 0.033 ML/s. These conditions were found to yield a QD density of about 400 QDs per $\mu m^2$ (see Ref. [23]). Finally, the substrate was capped at 480 °C by 100 nm GaAs to finish with a symmetric PC slab layer of $t$ = 200 nm thickness.

The series of PC samples fabricated have a triangular lattice with lattice constant $a$ ranging from 350 to 490 nm in 10 nm steps and air holes of radius $r$ = 0.3$a$. The PC nanocavities consist of a line of three missing air holes, i.e., the L3 donor type. By shifting the two adjacent air holes at the end of the L3-type defect, the $Q$ factor can be significantly increased. [1] Here,

L3 nanocavities with different air holes shift were used, i.e., $\Delta s = 0$ (no shift), $0.1a$ and $0.2a$. Their pattern was transferred into the GaAs slab by using electron-beam lithography and HI/Xe inductively coupled plasma etching. [25] In the next step, the underlying AlGaAs sacrificial layer was removed by HCl wet etching. The resulting structure confines light vertically by total internal reflection and in plane by its 2D photonic band gap. [26]

In order to compare the experimental data, the theoretical $Q$ factors ($Q_{theo}$) of the nanocavities were calculated by the three-dimensional (3D) finite-difference time-domain (FDTD) method, taking into account that the physical thickness $t$ of the slab is constant while the relative $t^*(a)$ thickness changes according to the lattice constant $a$. The grid cell size in the 3D FDTD calculation was meshed $\frac{\sqrt{3}}{16}a \times \frac{a}{10}$ in plane and $\frac{a}{10}$ out of the 2D-PC slab plane direction. For $\Delta s = 0$ $Q_{theo}$ ranges from 3400 to 5000, for $\Delta s = 0.1a$ from 9400 to 15 400, and for $\Delta s = 0.2a$ from 43 000 to 78 000, as the lattice constant changes from 490 nm ($t^* = 0.4a$) to 350 nm ($t^* = 0.6a$). $Q_{theo}$ becomes smaller as the lattice constant $a$ increases because the relative thickness of the slab $t^*$ becomes thinner and thus the optical confinement of the PC slab is less strong.

*Experimental work:* Q factor *measurement via the reflectance spectrum*

All measurements were carried out at room temperature (RT). Figure 1 shows a schematic of the experimental setup used for reflectance measurements. [20] A continuous wave tunable wavelength ($\lambda$) laser, linearly polarized at an angle defined as 0°, was used as a probe light source. For reflectance measurements, the light is focused by a microscope objective onto the top of the 2D-PC nanocavity of interest, whose cavity mode polarization – here its zeroth order mode – is aligned 45° relative to the incident light. The reflected light is collected by the same objective and passes a polarizer, set to 90° and focused on a photodiode detector. The orthogonal polarization configuration of the incident and detection light suppresses background reflectance which is mainly off the 2D-PC area surrounding the cavity. Thus, not only can artifacts in the spectrum that do not originate from the nanocavity be largely avoided but also the visibility of the nanocavity mode of interest is maximized.

Figure 2 illustrates examples of the reflection spectra and images, where measurements on and off the nanocavity's position are compared. Figure 2(a) shows the reflectance spectrum of a nanocavity with $a = 350$ nm and no shift, where the focus spot position in the 2D-PC nanocavity area is indicated by the inset of the figure. The spectrum has a sharp peak at 1274 nm indicating an increase in the reflection intensity at that specific wavelength. This wavelength is within the range of QD emission but we note that the resonance peak shape did not change appreciably with excitation power from 0.07 to 7 µW at RT. Consequently, any effect due to absorption and emission by QDs will not be considered here. Figure 2(b) is the IR image corresponding to Fig. 2(a) at the peak wavelength (1274 nm), where a bright reflected spot can be detected at the position of the nanocavity. In contrast, Fig. 2(c) shows the reflection spectrum of a position in the 2D-PC region off the nanocavity area, as shown in the inset and consequently, no characteristic peak can be seen in the spectrum. Figure 2(d) shows the corresponding IR image at 1274 nm, i.e., the peak wavelength of Fig. 2(a) but in Fig. 2(d) no characteristic reflection spot is detected. The comparison of the reflection spectra and images at positions on and off the nanocavity support the argument that the peak in the reflection spectrum in Fig. 2(a) certainly originates from the nanocavity's resonant mode.

Further support comes from the comparison of reflectance spectra with their conventional PL counterparts in Fig. 3. On the one hand, Fig. 3(a) shows the reflectance measurement results of nanocavities without shift ranging from $a = 350$ to 390 nm where a clear peak was observed for each of the five nanocavities (at 1274 nm, 1300 nm, 1324 nm, 1351 nm, and 1374 nm, respectively). Note, that the wavelength interval between the peaks is almost constant and therefore proportional to the linearly varying lattice constant and nanocavity size. Figure 3(b) on the other hand shows the spectra of the same nanocavities taken by conventional PL (colored solid lines). As a reference, the QD PL emission spectrum from an area of the substrate without PC structure taken at RT is also shown (the inset and the dashed line). [23] Figure 3(b) indicates that in PL the nanocavity peaks were detectable except for the sample with the greatest lattice constant ($a$ = 390 nm) and there is a good correspondence of the peak wavelengths between the PL and reflection data when compared to Fig. 3(a). In case of $a$ = 390 nm the resonant wavelength found by reflection agrees well to a linear fitting of the resonant wavelengths found by PL for $a$ = 350 to 380 nm. Figure 3 hints that QD PL emission couples to the nanocavity mode, if $a$ is smaller or equal to 380 nm and therefore its behavior is active whereas for $a$ greater than 390 nm a nanocavity's behavior is passive as it is detuned from any QD related emission (the inset and the dashed line). Thus the data clearly demonstrates that the reflectance method can appraise both active and passive nanocavities unlike PL methods that can probe only active nanocavities. We note that in Fig. 3(b) the nanocavity mode PL full width at half maximum (FWHM) linewidth is broad and almost constant because it is limited by the resolution of the spectrometer (1.8 nm). Although spectral resolution of the PL measurement can be improved, it is difficult to realize high resolution and high sensitivity simultaneously. By contrast, the linewidth appears much narrower in the reflectance measurement since the resolution is determined by the wavelength meter (0.3 pm).

*Analysis and discussion*

The experimental $Q$ factors ($Q_{exp}$) were evaluated from the linewidths of the reflectance peaks for cavities with $a$ ranging from 350 to 490 nm and different air hole shift $\Delta s$. To fit the experimental data, the following three contributions were taken into account. First the reflectance spectrum of the cavity mode was calculated in the weak excitation limit. The reflectance spectrum of passive and active cavities, assuming weak coupling between QDs and nanocavity mode for the latter, can be explicitly written as

$$R(\lambda) = \kappa(\lambda) \cdot \left| \frac{-\Gamma_c}{\Gamma_c - i(\omega - \omega_c)} \right|^2, \qquad (1)$$

where $\kappa(\lambda)$ represents the input/output coupling efficiency between the probe light beam vertical to the 2D-PC slab and the nanocavity mode, $\Gamma_c$ is the field decay rate of the nanocavity mode, and $\omega$ denotes the frequency of the probe light with $\omega_c$ being the resonant frequency of the nanocavity mode. Note, that Eq. (1) is a Lorentzian-type function. Second, asymmetry of the reflectance peak shoulders was included that account for the Fano effect. [27-29] Third, Fabry-Pérot resonances due to multilayer interference, as seen from the free spectral range, were added to the fit function by using the Airy method with transfer matrices. [30] Depending on the relative numbers of its $Q$ factor, spectral position, and intensity in the spectrum, there is a variety of possibilities for a Fabry-Pérot resonance to alter the nanocavity reflectance peak. The peak, as seen from the spectrum, can show asymmetry, a side peak, overlapping peaks, etc. Statistics, i.e., a large number of measurements, helps in a careful analysis. The occurring dispersion was not found to be significant. Finally, a constant background was added to the fit function to account for noise floor.

Figure 4 gives a survey of the $Q_{exp}$ [$\Delta s = 0$ (green triangles), $0.1a$ (yellow squares), and $0.2a$ (red stars)] and compares the data to the interpolated $Q_{theo}$ calculated by 3D FDTD and represented by the blue lines [$\Delta s = 0$ (solid), $0.1a$ (dotted), and $0.2a$ (dashed)]. The error bars indicate the standard deviation error. Good agreement between the theoretical prediction and the experimental data for $\Delta s = 0$ and $0.1a$ cavities can be seen. Also there is an overall agreement of the trend for $\Delta s = 0.2a$ cavities, where the $Q_{exp}$ factor is lower for longer wavelengths. This trend of $Q$ can be explained by the relative thickness $t^*(a)$ of the slab as mentioned before.

When comparing the spread of the experimental data there are two distinctive features: First, the detrimental influence of nonintrinsic loss components on $Q_{exp}$ becomes more severe with increasing $Q_{theo}$. The $Q_{exp}$ of L3 cavities with $\Delta s = 0$ and $0.1a$ are apparently robust against fabrication issues and detrimental environment influences that cause the nonintrinsic loss whereas in the case of $\Delta s = 0.2a$ the experimental data is well below the theoretical curve. It is also noted that the nonintrinsic loss seems to be less significant for smaller $a$ in the case of $\Delta s = 0.2a$. The reason for this is not entirely clear but is most probably related to the relative slab thickness. The highest $Q_{exp}$ reaches 58 000 for $a = 390$ nm, i.e., more than 90% of its $Q_{theo}$ of about 64 000.

The second distinctive feature in Fig. 4 appears over the wavelength range 1350–1440 nm. The scattering of $Q_{exp}$ is greater here than elsewhere, most clearly seen from the data of cavities without shift. In this wavelength range, the ratio of cavity-resonant peak maximum to background noise floor due to surface scattering in the reflectance spectra is very small. Moreover, within this range it was not possible to find the cavity related peak for some cavities with $\Delta s = 0.1a$ and $0.2a$. There are several reasons that could account for this behavior.

An effect of sample aging was experienced for some of the $\Delta s = 0.2a$ cavities, that yielded to sample degradation. [31] Besides this nonintrinsic effect, it was found that within the 1340–1480 nm range the strength of the nanocavity reflection signal can become as low as the background of Fabry-Pérot interference peaks which are damped by the cross polarization. This low signal-to-noise ratio (SNR) can make it challenging to distinguish the origin of a peak. Figure 5 shows the experimental SNR data (blue *y* scale on left), using the same symbols (solid blue) as Fig. 4 in order to separate between the different air hole shifts of the nanocavities, as a function of resonant wavelength.

Two reasons are considered for the low SNR observed in this wavelength range. First, water absorption in the free space beam path, in optical fibers, and adsorbed films at interfaces of optical components of the setup comes into play in this range. [32] By recording the reflection signal of a mirrorlike surface, sharp absorption lines are seen in the spectrum, shown as a reflection reference (solid blue line) in Fig. 6(a). Sharp absorption dips observed in this spectrum result in a decrease in the SNR and have the potential to overlap and hide narrow cavity lines completely. As indicated in Fig. 6(a) the sharp dips can also be seen in the reflection measurement data (black dots) in cross polarization, agreeing well with the dips in the reflection reference. In this example, the nanocavity ($\Delta s = 0.2a$ and $a = 390$ nm) reflection peak itself is free from dips as indicated by Fig. 5(b), where $Q_{exp}$ is about 58 000.

Furthermore, a second reason for the low SNR, is the weakness of the signal itself in that specific wavelength range. The optical length between the nanocavity and the underlying substrate surface is about the cavity's resonant wavelength, i.e., the PC slab center is situated in a node of the electric field of the incident light. Thus, the coupling of the light beam vertical to the slab and the nanocavity, considered as cavity mode and leaky mode, is diminished because their fields' overlap vanishes. The details, including an analytical model to obtain the reflectivity are given in the Appendix.

For an analysis, the maximum of the nanocavity's reflectivity in cross polarization denoted as peak reflectivity is calculated as a function of resonant wavelength. In Fig. 5, the peak reflectivity is plotted by the solid red line (red *y* scale on right), indicating that this parameter becomes minimal at around 1370 nm. This effect of a minimal reflectivity within a certain wavelength range in theory is supported by the experimental SNR data.

## *Conclusion*

In summary, we have demonstrated the measurement of the *Q* factor of nanocavities from the reflectance spectrum over a large wavelength range and pointed out that the reflectance method has the ability to probe both active and passive nanocavities. In particular, this feature allows for a direct comparison of the cavities' resonant wavelengths measured by reflectance and conventional PL and was found to be in good agreement within the wavelength resolution of the setup used. The experimental *Q* factors of nanocavities have been evaluated by the reflectance method and were confirmed to largely follow the theoretical trend as calculated by 3D FDTD due to the relative slab thickness. Additionally, the coupling to the nanocavity is affected by the coupling of the light beam vertical to the slab which can become minimal in a specific wavelength range.

The measurement technique affords both high resolution and high sensitivity and is believed to also be useful for probing higher order cavity, waveguide, nonlocalized, etc., modes. This will allow the measurement of CQED interaction phenomena and properties (e.g., QD absorption) in single, deterministically coupled emitter-nanocavity systems.

*Note added in proof.* we got aware of a recent publication of related work [33] during the revision process of this manuscript.

## *Acknowledgement*

We would like to thank Makoto Yamaguchi for helpful discussions and gratefully acknowledge support by Research Programs (Grant-in-Aid, COE, and Special Coordination Fund) for Scientific Research from the Ministry of Education, Culture, Sports, Science and Technology of Japan, and also by Core Research for Evolutional Science and Technology of the Japan Science and Technology Agency.

## *Appendix: Analytical model of the reflectivity*

In the following, an analytical model is proposed that describes the experimental observation of the nanocavity reflection in good approximation and is computationally less intense compared to 3D FDTD. This can be achieved by a modified transfer matrix method.

Figure 7 illustrates a cross section of the assumed system. Starting from the top (layer 0), a semi-infinite layer ($n_0$=1) forms the upper cladding of the PC slab. This slab, denoted as layer 1, shall have a thickness $d=t_1$ and an effective refractive index $n_1$. In the experiment, the sacrificial layer situated underneath the PC slab was removed by wet etching. Thus, it forms the lower cladding (layer 2, thickness $d=t_2$) and is assumed to have the same refractive index $n_0$ as the upper cladding. The bottom (layer 3) is formed by a semi-infinite substrate of refractive index $n_3$.

The electric field of the incident wave in *z* direction is given by *S* where its index *j* denotes the interface and the sign of the index the direction of *S*. Fresnel reflection and transmission is considered to be normal with respect to the *x-y* plane. A common method to treat multilayer systems is the transfer matrix method, [30] which is expanded to 4x4 matrices to account for the two orthogonal $x$ and $y$ polarizations (0° and 90°, respectively). The matrices can be viewed as 2x2 blocks of the pure polarizations along the main diagonal and off-diagonal elements as mixing terms. $\beta = \frac{n\omega}{c_0}$ is the propagation constant, where $c_0$ the speed of light in vacuum. The propagation matrix $T_p$ reads

$$T_p = \begin{pmatrix} \exp(-i\beta d) & 0 & 0 & 0 \\ 0 & \exp(i\beta d) & 0 & 0 \\ 0 & 0 & \exp(-i\beta d) & 0 \\ 0 & 0 & 0 & \exp(i\beta d) \end{pmatrix}. \quad (A.1)$$

The reflection and transmission at an interface $(j,j+1)$ is expressed by the transition matrix

$$T_{j,j+1} = \frac{1}{2n_{j+1}} \begin{pmatrix} (n_{j+1}+n_j) & (n_{j+1}-n_j) & 0 & 0 \\ (n_{j+1}-n_j) & (n_{j+1}+n_j) & 0 & 0 \\ 0 & 0 & (n_{j+1}+n_j) & (n_{j+1}-n_j) \\ 0 & 0 & (n_{j+1}-n_j) & (n_{j+1}+n_j) \end{pmatrix} \quad (A.2)$$

and the coupling to the nanocavity by

$$T_c = \frac{1}{i(\varpi - \varpi_0) + \frac{\varpi_0}{2Q_{loss}}} \begin{pmatrix} 1-\kappa_x^2 & -\kappa_x^2 & -\kappa_x\kappa_y & -\kappa_x\kappa_y \\ \kappa_x^2 & 1+\kappa_x^2 & \kappa_x\kappa_y & \kappa_x\kappa_y \\ -\kappa_x\kappa_y & -\kappa_x\kappa_y & 1-\kappa_y^2 & -\kappa_y^2 \\ \kappa_x\kappa_y & \kappa_x\kappa_y & \kappa_y^2 & 1+\kappa_y^2 \end{pmatrix}, \quad (A.3)$$

where

$$\kappa_{x/y} = \sqrt{\frac{\varpi_0}{2Q_{cav\ x/y}}} \quad (A.4)$$

describes the coupling constant between the light beam orthogonal to the PC slab and the nanocavity. Consequently, $\kappa_{x/y}$ comprises $Q_{cav}$, i.e., the vertical quality factor of the nanocavity and is about $10^4$. Using two different $Q_{cav}$ for $x$ and $y$ polarization can be used to include alignment deviations. The quantity $Q_{loss}$ corresponds here to the in plane $Q$ factor that represents loss components found by 3D FDTD and is on the order of $10^8$.

In order to evaluate the input-output relation of the entire system as shown in Fig. 7 the normalized initial condition (i.e., input of *x* polarized light from the top)

$$x \text{ polarization} \quad \begin{aligned} S_{x+0} &= 1 \\ S_{x-5} &= 0 \end{aligned}, \quad (A.5)$$

$$y \text{ polarization} \quad \begin{aligned} S_{y+0} &= 0 \\ S_{y-5} &= 0 \end{aligned}, \quad (A.6)$$

is put into the equation of the total transfer matrix of the system $T_s$

$$\begin{pmatrix} S_{x+5} \\ S_{x-5} \\ S_{y+5} \\ S_{y-5} \end{pmatrix} = T_{01} T_p(t_1/2) T_c T_p(t_1/2) T_{12} T_p(t_2) T_{23} \begin{pmatrix} S_{x+0} \\ S_{x-0} \\ S_{y+0} \\ S_{y-0} \end{pmatrix} = T_s \begin{pmatrix} S_{x+0} \\ S_{x-0} \\ S_{y+0} \\ S_{y-0} \end{pmatrix}. \quad (A.7)$$

Only *y* polarized output light is observed from the top ($S_{y-0}$). Thus, the system's transfer matrix is converted into a scattering matrix and the total reflectivity in cross polarization

$$R_{cross\ polarization}(\varpi) = \left| \frac{S_{y-0}(\varpi)}{S_{x+0}(\varpi)} \right|^2 \quad (A.8)$$

can be obtained. With this expression and the structural parameters of the nanocavity samples, the nanocavity resonant wavelength is kept fixed while the reflectivity is evaluated as a function of the wavelength $\lambda$. The maximum reflectivity or peak reflectivity over wavelength reproduces the red curve in Fig. 5. A similar analysis can be done for the total $Q$ factor that fluctuates within certain limits as in this model the material interfaces act as partial reflectors and thus can enhance or decrease the coupling to the nanocavity and thus, contribute or decrease the total $Q$.

*Figure captions*

Fig. 1

 (Color online) Schematic showing the method in principle and the experimental setup used. Incident light of a suitable light source is reflected from a 2D-PC nanocavity and sent to a wavelength meter for measurement or an IR camera for imaging. The cross polarization configuration suppresses unwanted background and maximizes the nanocavity mode.

Fig. 2

(Color online) (a) Reflection spectrum from the nanocavity (NC) with sample schematic (inset) and (b) IR image at the resonant wavelength (1274 nm) showing the appearance of a sharp peak strongly confined within the nanocavity area. (c) Reflection spectrum from the PC with the laser positioned as shown in the inset and (d) diffuse IR reflection at 1274 nm. Nanocavity is an L3 without shift ($\Delta s = 0$), $r = 0.3a$, and $a = 350$ nm.

Fig. 3

(Color online) (a) Reflectance spectra of L3 nanocavities without shift with various lattice constants $a$ (nm). (b) The corresponding mode PL spectra of the nanocavities. An offset was added for clarity. The spectral distribution of the QD intensity is overlaid as the dashed line. The top inset shows the QD ensemble PL intensity without PC in the range from 900 to 1400 nm (horizontal axis) in arbitrary units (vertical axis). In the inset the darkened area corresponds to the wavelength range of the main figure.

Fig. 4

(Color online) Experimental $Q$ factors for L3 nanocavities (with air hole shift $\Delta s = 0$, $\Delta s = 0.1a$, and $\Delta s = 0.2a$) compared to their theoretical trends calculated by 3D FDTD.

Fig. 5

(Color online) Comparison over wavelength of the experimental SNR data (left side, blue symbols) and the nanocavity peak reflectivity as calculated by the analytical model (right side, solid red line). The symbols are the same as used in Fig. 4 to distinguish the different air hole shifts.

Fig. 6

(Color online) (a) Comparison of the reference reflection of a mirrorlike surface (blue line) that shows sharp dips due to water absorption and the L3 nanocavity with $\Delta s = 0.2a$ and $a = 390$ nm (black dots). (b) Expansion of the wavelength range that shows that the nanocavity reflection peak data is free of such influence. The solid red line indicates the fitting function used for determining $Q_{exp.}$

Fig. 7

(Color online) Cross section of the analytical model for the calculation of the nanocavity reflectance.

*Figures*

W. Stumpf *et al.,* Fig. 1

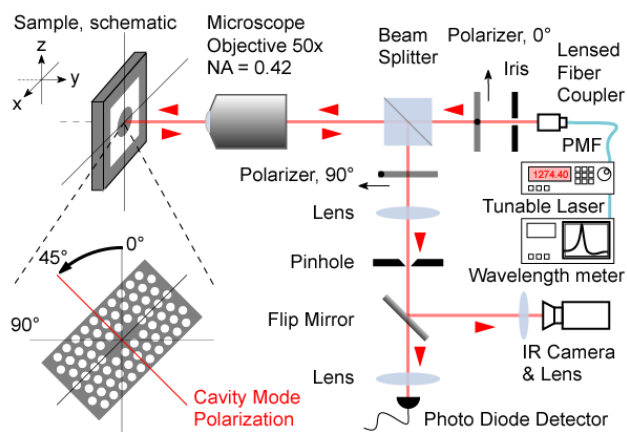

W. Stumpf *et al.,* Fig. 2

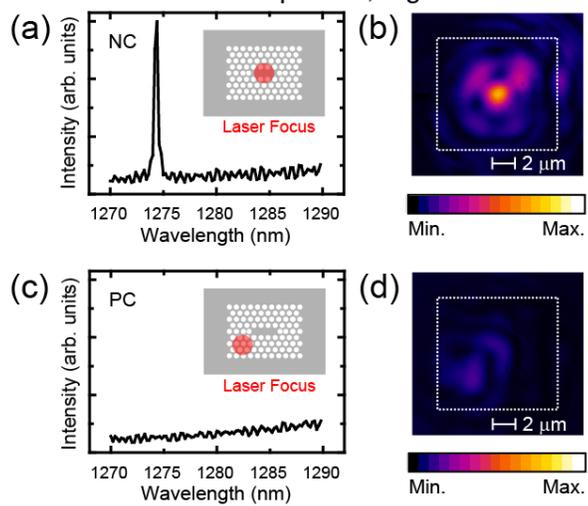

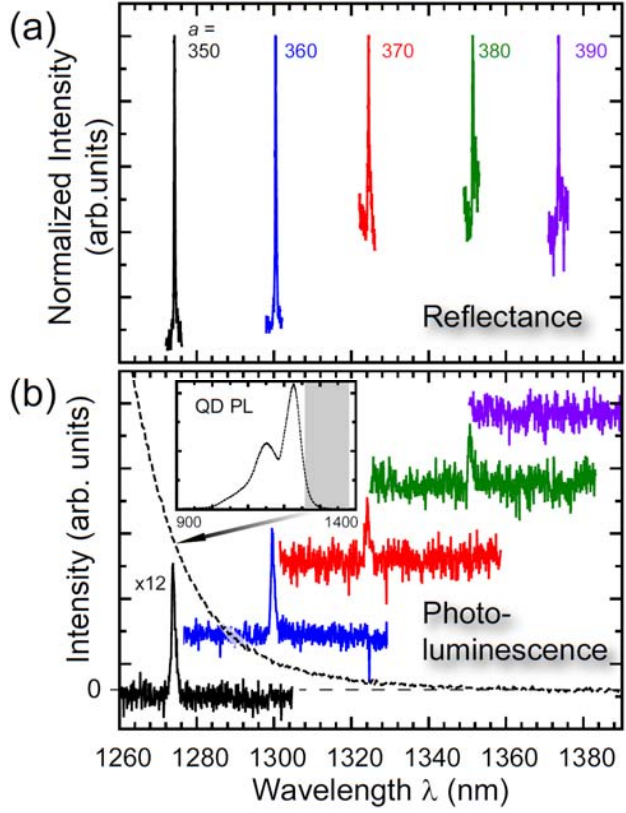

W. Stumpf et al., Fig. 3

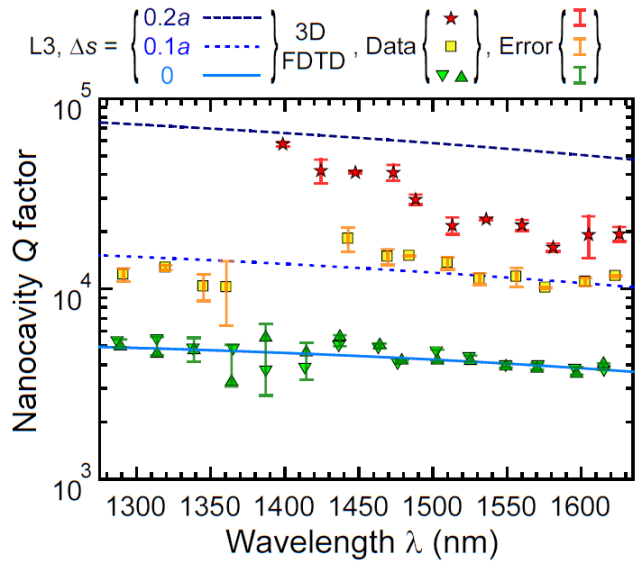

W. Stumpf et al., Fig. 4

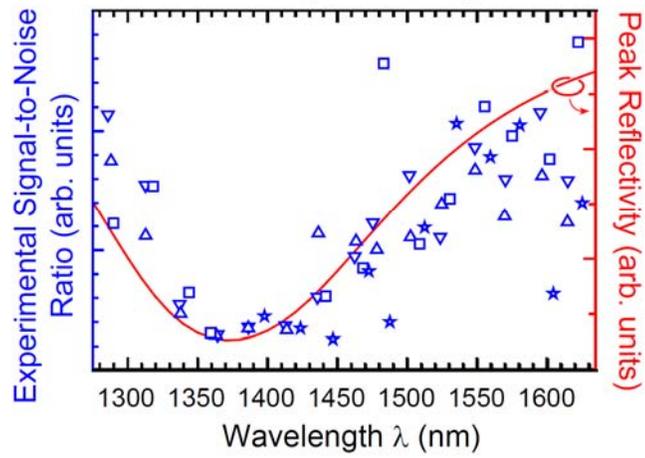

W. Stumpf et al., Fig. 5

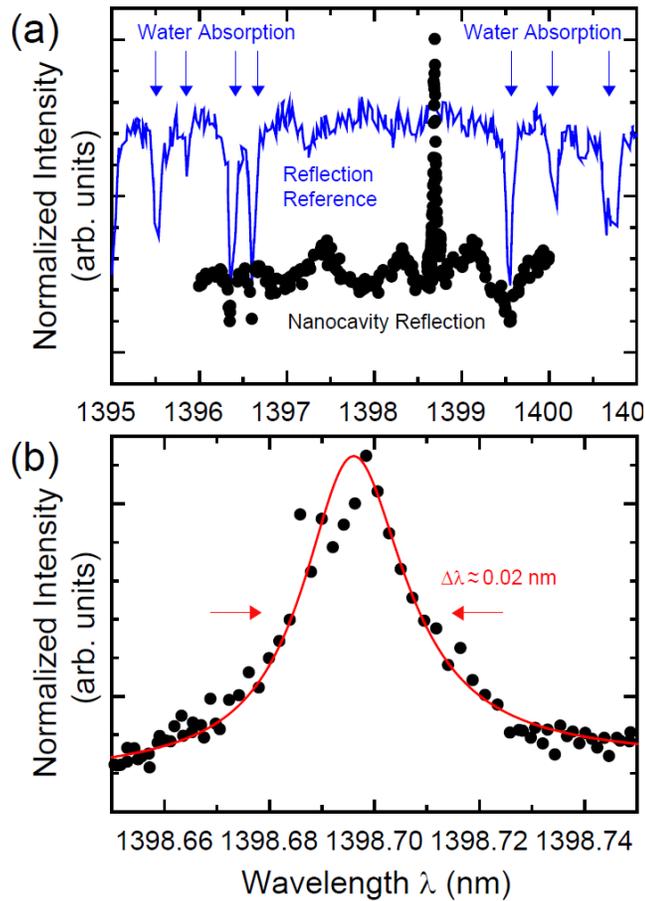

W. Stumpf et al., Fig. 6

W. Stumpf *et al.,* Fig. 7

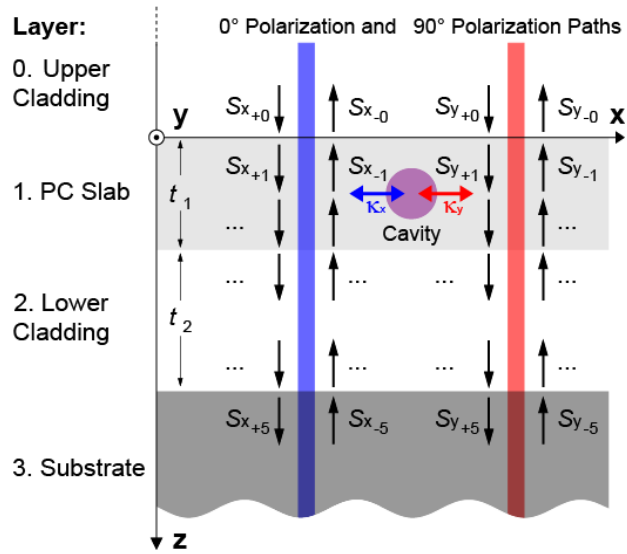
24